\documentclass[]{spie}  

\usepackage{amsmath,amsfonts,amssymb}
\usepackage{graphicx}
\usepackage[colorlinks=true, allcolors=blue]{hyperref}

\title{SKAO Observation Execution Tool: Designing for concurrent, responsive observations}

\author[a]{Viivi Pursiainen}
\author[a]{Stewart J. Williams}
\author[a]{Thaddeus Kenny}
\author[a]{Elizabeth S. Bartlett}
\author[a]{Andrew D. Biggs}
\author[a]{Brendan McCollam}
\author[b]{Danilo Acosta}
\author[b]{Sean Ellis}
\author[b]{Rupert Lung}
\affil[a]{UK Astronomy Technology Centre, Edinburgh, United Kingdom}
\affil[b]{CGI, London, United Kingdom}

\authorinfo{Further author information: (Send correspondence to V.P.)\\V.P.: E-mail: viivi.pursiainen@stfc.ac.uk}

\begin{document} 
\maketitle

\begin{abstract}
The SKA Observatory, currently in the construction phase, will have two of the world’s largest radio telescopes when completed in 2028. The scale of the project introduces unique challenges for the telescope software design and implementation at all levels, from user interfacing software down to the lower-level control of individual telescope elements. The Observation Execution Tool (OET) is part of the Observation Science Operations (OSO) suite of applications and is responsible for orchestrating the highest level of telescope control through the execution of telescope control scripts. One of the main challenges for the OET is creating a design that can robustly run concurrent observations on multiple subarrays while remaining responsive to the user. The Scaled Agile Framework (SAFe) development process followed by the SKA project also means the software should be allow to iterative implementation and easily accommodate new and changing requirements. This paper concentrates on the design decisions and challenges in the development of the OET, how we have solved some of the specific technical problems and details on how we remain flexible for future requirements.
\end{abstract}

\keywords{SKAO, OET, Observation Execution Tool}

\section{INTRODUCTION}

The SKA Observatory (SKAO) will comprise two radio telescopes: SKA-Mid, with 197 dishes covering frequencies 350\,MHz-15.4\,GHz, and SKA-Low, featuring 512 stations of 256 antennas operating within the 50-350\,MHz range. To manage proposals and observations throughout their lifecycle, the Observatory Science Operations (OSO) application suite is being developed. The OSO suite comprises a set of applications dedicated to supporting each stage of the lifecycle, from proposal preparation and assessment through to observation design, scheduling, and execution \cite{Kenny24}.

The Observation Execution Tool (OET), as part of the OSO suite, is responsible for managing the execution of Scheduling Blocks (SB) on telescope subarrays. Each SB is a self-contained definition of the abstract telescope resources, configurations, and timing requirements to perform an astronomical observation. The SKA telescopes can be divided into up to 16 subarrays per telescope, each capable of performing an independent observation. Scheduling Blocks are the input to Python-based Observing Scripts, which translate the SB at run-time into a sequence of configuration and control instructions sent to the SKA Telescope Monitoring and Control system to achieve the observation defined in the SB. The OET runs up to 16 Observing Scripts in parallel, each script performing its own independent observation on a dedicated telescope subarray. 

The main drivers of the OET design are the requirements of robustness and responsiveness when executing simultaneous observations across multiple subarrays. A further challenge is supporting the iterative development of Observing Scripts in a Kubernetes-based container orchestration environment, where the OET deployment is long-lived and fixed yet the Observing Scripts that the OET runs evolve and are developed on a rapid timescale. 

In this paper, we describe several features of the OET's architecture and design that address these scalability and performance requirements. This will include an overview of the event-driven Python multiprocessing architecture and our solution to dynamic Observing Script support using git as the script development platform. 

\section{Requirements}
The OET is designed to meet a set of requirements captured in the OSO Software Architecture Document (SAD)\cite{SKASAD}. The SAD outlines an initial design for all OSO applications centred around a `shared data' architecture, where applications access data via a common database rather than via application-to-application data transfer. The OET architecture proposed in the SAD and refined over the course of implementation aims to satisfy the following key performance requirements on concurrent and responsive observation execution:

\begin{enumerate}
    \item The OET must be robust, as SB-driven observations are the primary observing mode for SKA and they depend entirely on the OET.
    \item As a system, the OET must support simultaneous observations on up to 16 subarrays. Observations on subarrays may be independent, where failure on one subarray does not affect other subarrays, or they may be linked, where failure on one subarray halts observing on the subarray observations that depend on it.
    \item The OET must support the execution of time-critical Target of Opportunity (TOO) observations. These observations require an immediate response from the OET, with minimal latency from receiving the TOO trigger to aborting the current observation and executing the requested TOO observation.
    \item OET should support dynamic and fast changing Observing Scripts. This is complicated, as the Observing Scripts could be developed after OET delivery, use third-party libraries, and be written by users rather than professional software developers. 
\end{enumerate}

The current deployment configuration plan will address requirements 1 and 2 by deploying one OET per subarray. To guarantee robustness and prevent one failure cascading to other subarrays, each subarray is controlled by its own instance of the OET. This isolation naturally prevents failures on one subarray from affecting other observations. A dedicated monitoring component is responsible for detecting any issues with the OET or Observing Script, allowing linked observations to be halted if a problem arises. This monitoring component adds support for compound observations spanning multiple subarrays, where all observations must succeed for the aggregated data to be useful.

To address requirement 3, an early prototype of the OET application demonstrated that an approach using Python multiprocessing could help in meeting the latency requirement\cite{Williams16}. A first implementation for the OET application has been created based on the prototype and is described in §\ref{sec:architecture}.

While OET development is guided by requirements and design outlined in the SAD, several elements have evolved due to new requirements discovered during construction. These changes are easily accommodated in SKAO's development cycle using Scaled Agile Framework (SAFe) where the focus is incremental development to allow for evolving requirements\cite{Klaassen20}. One such requirement is requirement 4 for dynamic Observing Scripts. To allow flexible and maintainable Observing Script development, OET provides the option of retrieving scripts from git and running them in a custom Python environment. This approach has been described in more detail in §\ref{sec:dynamic observing scripts}.

\section{Architecture}
\label{sec:architecture}
The OET is based on Domain-Driven Design\cite{Evans04} (DDD) concepts, with a split between a user-focused SB execution domain and a more technical script execution domain. To summarise OET operation, a telescope operator request to run an SB becomes an event received by the script execution domain, which triggers the running of the Observing Script referenced by the SB in a new Python subprocess. As the Observing Script runs, it emits events (`configuring telescope', `performing scan', etc.) which can be received, presented, and/or recorded by various other SKAO software components (user interfaces, project tracking, etc.). The OET monitors and reports the subprocess status and can take action if necessary, such as terminating the subprocess running an unresponsive script.

Each subarray can only execute one observation at a time, meaning only one Observing Script can control the subarray at any moment. However, to facilitate rapid responses to TOO events, the OET allows multiple observing scripts to be prepared in advance. During this preparation stage, which can run concurrently for multiple scripts, the OET loads the script into memory, runs the initialisation function if necessary (for example to initialise variables or download data required by the main script) and prepares the execution environment. The specific steps in initialising the script depend on its origin and will be discussed further in §\ref{sec:dynamic observing scripts}.

\begin{figure} [ht]
   \begin{center}
   \includegraphics[height=10cm]{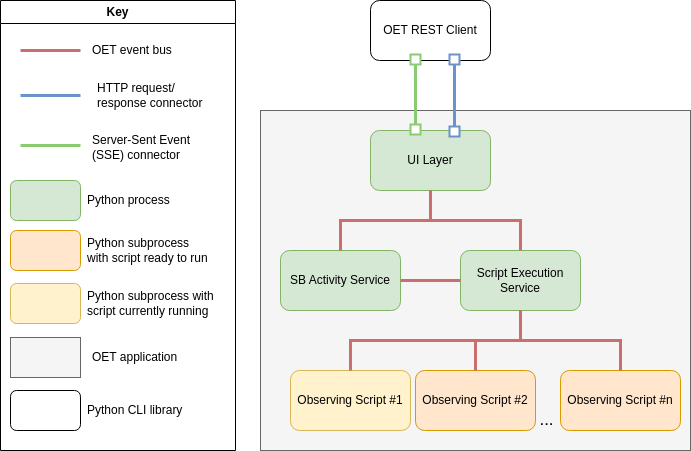}
   \end{center}
   \caption[OET internal multiprocessing architecture]
   { \label{fig:multiprocessing} OET internal multiprocessing architecture }
   \end{figure}

The OET architecture is based on event-driven multiprocessing. The application is divided into three services providing varying levels of abstraction: the Script Execution service, Activity service and the API layer. Each independent part of the application runs in its own process receiving and sending commands through the OET's event bus as displayed in Figure \ref{fig:multiprocessing}. This decoupled architecture is designed to maintain responsiveness in all levels of operation so that OET is able to respond to any events. Furthermore, within the Script Execution service each Observing Script is run in a Python subprocess so that the application can remain responsive to the user.

Details of the key responsibilities and interfaces of each service are given in the sections below.

\subsection{SB Activity service} 
When users and telescope operators initiate an observation, they don't want to think about specific scripts or parameters - they simply want to "run an SB". The SB Activity service, part of the user-focused SB domain, acts as a translation layer between the user and the Script Execution service. Instead of requiring users to specify Observing Scripts with precise parameters, such as SB IDs, subarray IDs, and script parameters, the user can request that an activity be performed on a specific SB. The SB Activity service then translates this request into events for the Script Execution Service, which will execute the script associated with the requested SB.

\subsection{Script Execution Service} 
The Script Execution Service, part of the script execution domain, is the core of the application and controls low-level script execution and management. It offers commands to prepare, execute and abort the execution of scripts and keeps a record of each script's execution history, including any stack trace, should errors be raised in the preparation or running of the script. In the script execution domain, the application is focused on Python script execution and has no logic or domain knowledge of astronomical observations and is not aware that the scripts being executed are sending commands to a telescope.

Executing a script is a two-phase process, split into 'prepare' and 'run' phases. Whenever a request to prepare a script for execution is received by the Script Execution Service, it triggers the launch of a bootstrap component which executes in a new Python subprocess. This bootstrap component handles retrieval and loading of the requested Observing Script, sets up the execution environment for the script and then awaits further signals from the Script Execution Service to invoke functions on the loaded script. The design takes advantage of Python's multiprocessing capabilities, allowing multiple scripts to be prepared simultaneously, creating a more streamlined and efficient workflow for Observing Scripts that require longer setups and initialisation.

\subsection{UI layer} 
The public interface for Script Execution and SB Activity application services lies in code held in the UI layer. The UI layer holds an API module, which presents methods of the Script Execution and SB Activity application services as REST resources accessible via HTTP. Additionally, the UI layer publishes OET events as a Server-Side Events (SSE) data stream, providing users with real-time insights into actions taken by the OET and Observing Scripts as they run.

\section{Dynamic Observing Scripts}
\label{sec:dynamic observing scripts}
The Observing Scripts executed by the OET are standard Python scripts that can import and utilize various Python libraries. The OSO Scripting library is one such library, offering the user a set of high-level functions to translate SB elements into commands for the Telescope Management and Control devices.

A significant challenge for the OET lies in managing the rapid evolution of Observing Scripts. While the OET itself is well-tested, the scripts it runs will likely undergo frequent updates, will not be as well tested, and will require a fast cycle for deployment. As scripts evolve, they may require new or updated library dependencies not present in the deployed OET environment. However, the OET's Python multiprocessing approach offers an advantage: scripts running in subprocesses also operate in independent Python interpreters, allowing different versions of scripts to run without OET redeployment, and for scripts to import different versions of their dependencies, all decoupled from the base OET environment.

\begin{figure} [ht]
   \begin{center}
   \includegraphics[height=9.5cm]{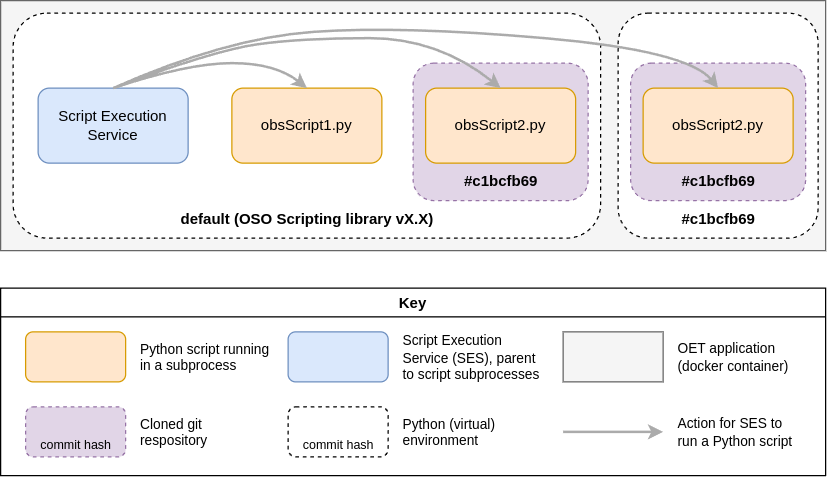}
   \end{center}
   \caption[Observing Script Python environment options]
   { \label{fig:obsScriptEnvs} Observing Script Python environment options }
   \end{figure}

To cater to scripts of all levels of maturity, OET supports tthree different approaches to Observin Script execution. These approaches are visualised in figure \ref{fig:obsScriptEnvs} and described below.

\textbf{1. From the OET filesystem:} This is for stable Observing Scripts with stable dependencies. They apply to general observing scenarios and have been thoroughly tested. Any Python or other dependencies are part of the default OET environment and don’t require additional packages to be installed.

\textbf{2. From a git repository without dynamic dependencies:} For Observing Scripts that are frequently changing but are running with stable dependencies. No packages need to be installed or upgraded to run the script but the script either does not exist in the filesystem of the deployed OET or is a newer version of an existing script.

\textbf{3. From a git repository with dynamic dependencies:} For Observing Scripts with frequently changing content and dependencies. Execution of these types of scripts requires OET to create a new Python environment to run the script in and so will have an increased time overhead but brings flexibility to the script dependency versions. This can be useful, for example, when the scripting library commands are updated but the updates have not yet made their way to the stable OET execution environment.

How this has been realised in the OET implementation, is by taking advantage of the existing multiprocessing design and combining it with Python’s virtual environments. As mentioned in the §\ref{sec:architecture}, each Observing Script is prepared and run in a child process of the Script Executor. This means that any loading, whether that is from filesystem or git, can be done in the background while leaving the control to the user to execute any further commands. This is especially useful when a dynamic execution environment is required, since installation of Python packages takes significantly longer than simply loading a script from the filesystem. 

In cases of executing scripts of types 2 and 3, OET will retrieve the script based on a given git repository by cloning the repository and loading the script from the cloned repo. The user can additionally specify a commit or branch if the script is not present in the latest main branch. For scripts of type 3, there is an extra step where on top of cloning the repository, OET creates a virtual environment where the script process installs the new requirements based on the git project’s pyproject.toml or requirements.txt files. These installed packages are then added to the subprocess' Python path. Since the environments are stored in the Script Executor’s main process, they can be reused by any future scripts when required allowing the script process to skip the dependency installation step.

These environments can only be reused by a different script if the required packages and the package versions are identical. As a first implementation, unique environments are identified through git commit hashes and saved in the OET application memory. We recognise that identifying identical environments through git commit hashes is not the most efficient solution and we outline some possible improvements to this approach in §\ref{sec:conclusion}.

\section{Conclusion and Future Work}
\label{sec:conclusion}

The initial implementation of the Activity layer and the use of dynamic Observing Scripts has been successful. In a simulated telescope environment (as there is no general access to hardware yet), OET is able to run an automated Scheduling Block driven observations where the SB is retrieved from the OSO Data Archive (ODA) by OET, the Observing Script referenced by the SB is loaded, either from filesystem or git, and executed sending the correct commands to the simulated Telescope Management and Control devices.

In the current implementation, each script process is terminated when the script has been executed. This means that if the same script is to be run again, it will need to be reloaded and prepared. In the future it could be more efficient to load the script in the child process and instead of terminating the process at the end of script execution, the process would be left to wait for another run command. This would be especially useful for when OET is required to run a list of observations that all follow the format of a standardised and stable Observing Script.

There are also some opportunities for more efficient script environment management. The current logic for virtual environment reuse is fairly basic. Whenever a script is requested to run from its own python environment, OET checks if an environment has already been created for the commit hash of the given repository. This is to speed up the environment creation process but it could still result in multiple duplicate environments as the dependencies do not necessarily change on every commit. A way to  do this more efficiently could for example include relating each environment with a hash of the local pip dependencies and identifying identical environments that way. This would however be more time consuming than the current method and the trade off between the two approaches would need to be evaluated.

While there are still some interface and performance improvements that can be made, the core architecture and design are yielding promising results at the current stage of implementation and are proving to provide the flexibility and responsiveness required by SKAO.

\bibliography{report} 

\begin{thebibliography}{1}

\bibitem{Kenny24}
{Kenny}, T., {Williams}, S.~J., {Pursiainen}, V., {Bartlett}, E.~S., {McCollam}, B., {Biggs}, A.~D., {Ellis}, S., {Lung}, R., and {Acosta}, D., ``{Development of the observatory software for the SKA},'' in [{\em Software and Cyberinfrastructure for Astronomy VIII}{\nolinebreak\hspace{0.1em}]},  {\em Society of Photo-Optical Instrumentation Engineers (SPIE) Conference Series} (2024).

\bibitem{SKASAD}
Nicol, M., Williams, S.~J., Reed, S., Valame, S., Canzari, M., Knapic, C., and Jerse, G., ``{OSO Software Architecture Document}.''

\bibitem{Williams16}
Williams, S.~J., Bridger, A., Chaudhuri, S.~R., and Trivedi, V., ``{The SKA observation control system},'' in [{\em Software and Cyberinfrastructure for Astronomy IV}{\nolinebreak\hspace{0.1em}]},  Chiozzi, G. and Guzman, J.~C., eds.,  {\bf 9913},  99132L, International Society for Optics and Photonics, SPIE (2016).

\bibitem{Klaassen20}
Klaassen, P.~D., Williams, S.~J., Nicol, M., Alberti, V., Bridger, A., Chrysostomou, A., Valame, S., Bartlett, E.~S., Canzari, M., Deolalikar, A., Lightfoot, J., McDermott, A., Pursiainen, V., Ribero, H., and Sabater, J., ``{Observatory science operations tool development for the SKA within a scaled agile framework},'' in [{\em Observatory Operations: Strategies, Processes, and Systems VIII}{\nolinebreak\hspace{0.1em}]},  Adler, D.~S., Seaman, R.~L., and Benn, C.~R., eds.,  {\bf 11449},  114490Z, International Society for Optics and Photonics, SPIE (2020).

\bibitem{Evans04}
Evans, E.,  [{\em Domain-Driven Design: Tackling Complexity in the Heart of Software}{\nolinebreak\hspace{0.1em}]}, Addison-Wesley (2004).

\end{thebibliography}
\bibliographystyle{spiebib} 

\end{document}